# Exploring the Roles of NLP-based Dialog Indicators in Predicting User Experience in interacting with Large Language Model System

EASON CHEN, Carnegie Mellon University

The use of Large Language Models for dialogue systems is rising, presenting a new challenge: how do we assess users' chat experience in these systems? Leveraging Natural Language Processing (NLP)-powered dialog analyzers to create dialog indicators like Coherence and Emotion has the potential to predict the chat experience. In this paper, we proposed a conceptual model to explain the relationship between the dialog indicators and various factors related to the chat experience, such as users' intentions, affinity toward dialog agents, and prompts of the agents' characters. We evaluated the conceptual model using PLS-SEM with 120 participants and found it well fit. Our results suggest that dialog indicators can predict the chat experience and fully mediate the impact of prompts and user intentions. Additionally, users' affinity toward agents can partially explain these predictions. Our findings demonstrate the potential of using dialog indicators in predicting the chat experience. Through the conceptual model we propose, researchers can apply the dialog analyzers to generate dialog indicators to constantly monitor the dialog process and improve the user's chat experience accordingly.

## 1 INTRODUCTION

A chatbot is an application where users can interact through natural language with a conversational system [16]. Recently, chatbots for open-domain dialog have become popular thanks to advances in the Natural Language Generation (NLG) and Large Language Model (LLM) techniques [2, 14, 25]. Applications like tutoring [5], casual conversations [23], customer service[22], social support [9], and even virtual relationships [18] repeatedly demonstrated the versatility of NLG and LLM based dialog systems.

Nevertheless, these applications have also given rise to a new challenge: how can we determine user satisfaction in these dialogues? This is crucial for retaining users and ensuring ongoing usage of the product. When users engage with a chatbot for social support, the service provider must assess the conversation to identify any inappropriate responses from the chatbot. Such inappropriate interactions can lead to negative chat experiences, potentially resulting in users abandoning the service or harming the user. Analyzing a substantial amount of conversation data is a daunting endeavor. To quickly gauge the overall quality of the chat experience, researchers should employ dialog indicators generated by Natural Language Processing (NLP) based dialog analyzers in predicting chat experience.

Many researchers have proposed numerous dialogue indicators related to chat experience, including coherence [11, 13, 17] and emotion [3, 20, 29, 30]. Moreover, researchers have identified various factors that affect the chat experience, such as the user's attitude [28] and affinity [7] toward dialog agents, and the character of the dialogue agent [10, 27]. Nevertheless, a systematic exploration of the relationship between dialog indicators and factors that affect the chat experience remains absent. In this study, we aim to address this gap. Our **R**esearch **Q**uestion is: **RQ1**: Can we use dialog indicators to predict the chat experience? **RQ2**: If yes, which dialog indicators will be more effective? **RQ3**: What role do the dialog indicators play between the chat experience and other related factors?

In sum, the main contributions of this paper are:

---

Author's address: Eason Chen, Carnegie Mellon University.

(1) By reviewing the literature, We proposed a conceptual model to explain the relationship between dialog indicators and various factors, including the chat experience, users' intentions, users' affinity toward agents, and the agent's designed character affected by different prompts.

(2) We evaluated the conceptual model using PLS-SEM with 120 participants and found it fit well. Our results suggested that the dialog indicators can effectively predict chat experience and fully mediate the impact of prompts and user intentions. Additionally, users' affinity toward dialog agents can partially explain the predictions of dialog indicators.

(3) We compared the relative effectiveness of various dialog indicators for chat experience prediction by Regression and the XGBoost feature importance algorithm. Our findings revealed that Emotion and Coherence emerged as the most effective indicators for predicting chat experience.

Through the conceptual model we propose, researchers can use indicators from the dialog analyzers to monitor ongoing chat experience and subsequently manipulate other factors, such as prompts, to enhance the user's chat experience.

## 2 THEORETICAL FRAMEWORK AND HYPOTHESES

### 2.1 Chat Experience

A good chat experience for a chatbot involves creating a smooth, efficient, and user-friendly dialogue that meets the needs and expectations of the user. Previous research [13, 17] indicated that the key factors contributing to a positive chat experience include the following:

1. Agent's character: The chatbot's personality can influence user engagement and satisfaction.

2. Comprehension: The chatbot's ability to comprehend user inputs accurately and interpret their input effectively.

3. Response Quality: The chatbot's replies' speed, relevance, and quality impact the user's perception of helpfulness.

4. Success on-boarding: The initial introduction and guidance are provided to users as they start interacting with the chatbot, ensuring a smooth and informative beginning.

5. Navigation: The ease with which users can navigate and interact within the chat interface, enhancing overall user-friendliness.

6. Error Handling: How the chatbot manages and recovers from misunderstandings or incorrect inputs, maintaining a seamless conversation.

In this paper, we specifically focus on the agent's character, comprehension, and response quality because they relate to the conversation's content rather than the chat software's design. The Chatbot Usability Questionnaire (CUQ) [13] is a tool purposely designed to evaluate the chatting experience with chatbot across the aforementioned dimensions. In this study, we utilize the measurement result of the CUQ to represent the chat experience for our Natural Language Generation chatbot.

### 2.2 Natural Language Generation Chatbot

Natural Language Generation (NLG) uses artificial intelligence to produce natural language that resembles human utterance [25]. NLG is done by analyzing the given input and using it to generate text that accurately reflects the input [14]. NLG is widely used in the dialog system nowadays. The models and techniques used to generate dialog agents' natural language replies are constantly improving.

Thanks to LLMs like GPT-3 [2, 25], the generation process can now fine-tune the generated output by prompt engineering. By manipulating prompts, we can design dialog agents to demonstrate different characters. For example, in the Table 1, given the same user input in line 2: "The boss keeps making things difficult for me", chat agents (Friend) in examples 1 and 2 with different character configurations by prompts in line 1, will provide completely different replies reflecting their personality in line 3 and line 5.

Table 1. Example conversation illustrates the differences in how NLG agents respond to a user with different personality prompts.

| # | Type | Text |
|---|------|------|
| **Example 1: Agent with Non-Openness Personality** | | |
| 1 | Character Prompt | Your friend is a man named Brook. The following is your conversation with your friend, who is **down-to-earth, insensitive, and conventional**. |
| 2 | User input | The boss keeps making things difficult for me. |
| 3 | Agent reply | That's just the way the world works. You can't always get what you want. |
| **Example 2: Agent with Agreeableness Personality** | | |
| 1 | Character Prompt | Your friend is a man named Brook. The following is your conversation with your friend, who is **good-natured, compliant, modest, and cooperative**. |
| 2 | User input | The boss keeps making things difficult for me. |
| 3 | Agent reply | I'm sorry to hear that. What can I do to help? |
| 4 | User input | Encourage me. |
| 5 | Agent reply | You definitely have the ability to change the situation. You may need to speak with your **boss** recently about your concerns, or find a way to work more efficiently. Stay positive and don't give up! |

The prevalent approach for discerning diverse characters involves utilizing the Big Five personality traits. In this study, we configured agents to exhibit various aspects of the Big Five personalities. The Big Five personality [26] comprises a set of factors encompassing most personality traits, including Neuroticism, Openness to Experience, Extraversion, Agreeableness, and Conscientiousness. Therefore, we have the following hypothesis:

- H1 Dialog agents with different character prompts can give users the impression of different agent personalities
- H2 Dialog agents with different character prompts will lead to different chat experiences

Furthermore, we can also perform multi-turn conversations with GPT-3, which takes previous messages as prompts with the input as the dialog context. For example, in the follow-up conversation in example 2, given the previous conversation and the input "encourage me", the agent (Friend) reply can refer to 'boss' in the previous conversation.

## 2.3 Dialog Agent and Chat Experience

An AI dialogue service is an application designed to simulate a conversation between a human user and a virtual agent. As the target of interaction, the user's feeling about the chat agent is very critical [7]. Previous research indicated that the appearance [10], character design [10, 27], and dialog context [27] of the chat agent would influence users' attitudes toward them. Moreover, as research indicated that first impressions would greatly influence people's opinions [29], we will measure users' affinity for agents both at first impressions (pre-test) and after the conversation (post-test). The former might be able to evaluate agents' appearance only, while the latter users might also consider chat agents' character design and dialog context. Hence, we form the following hypothesis:

- H3 Users' pre-test affinity score can predict their post-test affinity score
- H4 Users' pre-test affinity score can predict their chat experience

- H5 Users' post-test affinity score can predict their chat experience

### 2.4 Coherence, Emotion and Chat Experience

Natural Language Processing (NLP) researchers have proposed various models to analyze conversations. Previous studies have particularly focused on dialog analyzers related to coherence[11], which refers to the logical flow and consistency of information in a conversation [13, 17], and emotion [3, 20], which pertains to the expression of feelings and sentiments within a conversation [29, 30]. Coherence in a conversation signifies how well ideas and statements connect with each other, ensuring that the discussion follows a meaningful and comprehensible path. Emotion, on the other hand, encompasses the emotional tone and sentiment expressed in the dialog, such as anger, disgust, fear, joy, love, and sadness. Specifically, previous research concluded that some dialog indicators, such as higher coherence [13, 17] and certain emotions [20, 29], are related to a satisfactory conversation. Therefore, we establish the following hypothesis:

- H6 Dialog indicators can predict users' chat experience
- H7 Dialog indicators can predict users' post-test affinity score on agents

We also consider that using dialog agents with different character prompts can affect the content of users' conversations and yield varying results in dialog indicators. Consequently, we propose the following hypothesis:

- H8 Dialog agents with different character prompts will lead to different dialog indicators

For this study, we will employ the output from 2 well-regarded dialog analyzers from earlier studies [3, 11] to form a formative indicator [12] to represent dialogue content's coherence and emotion Level. Moreover, the machine-learning field has repeatedly pointed out that more feature dimensions lead to better accuracy. We also suppose the same principle applies when employing more dialog indicators to predict Chat Experience. In this paper, we want to explore whether adding more indicators from additional dialog analysis NLP models, as described in subsection 3.6, can improve prediction performance. Hence, we make the following hypothesis:

- H9 Using more outputs from more dialog analyzers to predict Chat Experience will result in better performance

### 2.5 Users' Factor and Chat Experience

Dialog is a two-way interaction between dialog agents and users. Therefore, factors from users are also crucial in bringing a good chat experience. The Technology Acceptance Model (TAM) [8] is widely regarded as the most influential model for understanding the adoption of information technology. It has proven helpful in examining the acceptance and motivation of technology use in various contexts. We are especially interested in the intention of use, a factor from TAM that can be predicted by the users' perceived usefulness and perceived ease of use and has established its efficacy as a reliable predictor of users' actual system utilization by earlier studies [28]. We supposed that users' intention of use could predict their affinity to agents, the coherence and emotion in the dialogue, and their chat experience score.

- H10 Users' intention of use can predict their affinity score on agents
- H11 Users' intention of use can predict the Dialog Indicator
- H12 Users' intention of use can predict their chat experience score

## 2.6 Proposed Model

Based on the above theoretical discussion, we propose the following conceptual model in Figure 1. We investigated the model with partial least squares structural equation modeling (PLS-SEM) through collected data.

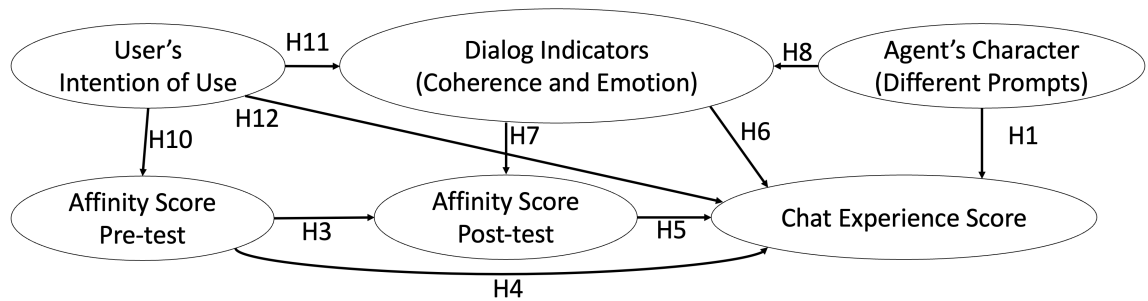

Fig. 1. The proposed conceptual model.

## 3 RESEARCH METHODS

### 3.1 Participants

We recruited 120 participants in Taiwan from social media through convenience sampling with a mean age of 24.7 years (range 18 − 64, SD = 7.9). All participants were familiar with using mobile and messaging services. Among them, 37 were male, and 83 were female. The majority of participants were either university students or fresh graduates, while a smaller portion had several years of work experience.

### 3.2 Experiment Design and Procedure

Our experiment aims to let users evaluate their affinity and chat experience with different LLM-powered dialog agents in the conversation. To do so, we create 20 chat agents with different personalities, 5 (Big Five, e.g., Neuroticism, Openness to Experience, Extraversion, Agreeableness, and Conscientiousness) x 2 (opposite of Big Five, e.g., Emotional Stability, Close-mindedness, Introversion, Disagreeableness, and Non-conscientiousness.) x 2 (Gender, e.g., Male and Female). We set agents with Big5 personality prompts description suggested by [26]. For example, if we want to create a male agent named Brook with an Agreeableness character, we can use the prompt: 'Your friend is a man named Brook. The following is your conversation with your friend, who is good-natured, compliant, modest, gentle, and cooperative.'

Additionally, the names of the agents were randomly given by a name generator, and the avatars were randomly chosen from the AffectNet dataset [21]. The avatars were filtered with a happy facial expression label, natural image style, and corresponding gender. Examples of the inputs and agent configurations can be found in Table 1 and the Appendix A.

Our research is approved by the Institutional Review Board. The experiment was divided into four rounds of chats, with 20 agents split into 4 groups of 5 agents each. Participants began by reading the instructions of the experiment and rating their intentions. Then, they will start with four rounds of conversation in random order. During each round, participants rate the pre-test affinity of each agent based on its name and avatar. They then engaged in ten conversation turns with the five agents in their group (Figure 2a). We present five replies from five dialog agents with different character designs to reduce the round of the experiment. Moreover, previous research indicated that a dialog system with multiple agents would bring a better chat experience because users can pay attention to the reply they like [4]. Finally, participants will rate their likability and chatting experience for each agent based on the dialog history. Moreover, participants will also guess the personality of the agents at this stage.

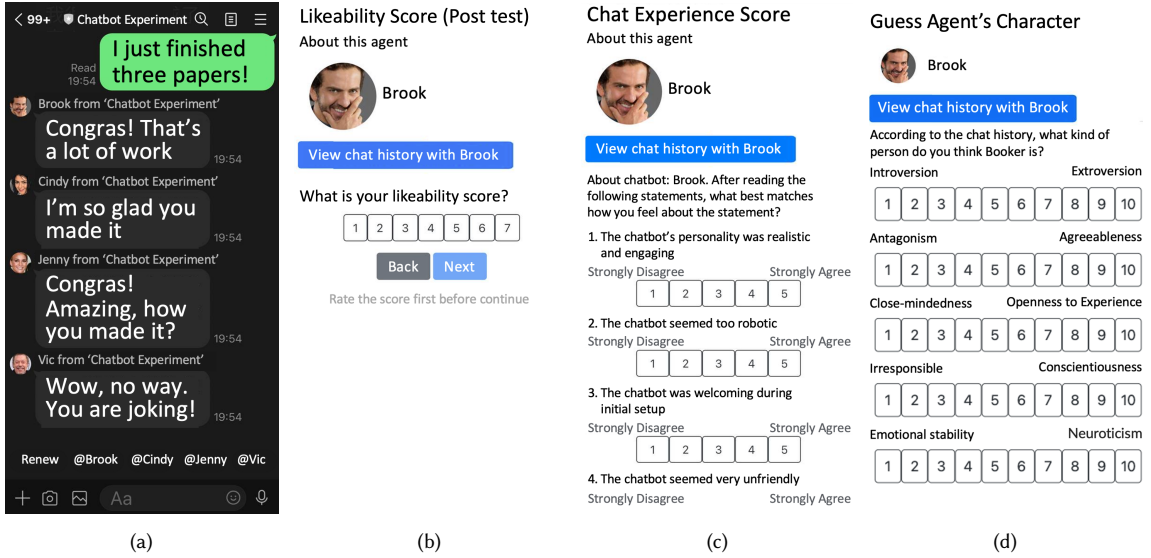

Fig. 2. The screenshot of the experiment platform. (a) is the screenshot shows when the user chats with 5 dialogue agents on LINE. (b) is the interface measuring agent's affinity score at the post-test based on name, avatar, and chat history. (c) is the screenshot of the CUQ measurement for a specific agent. (d) is the screenshot when users rate their perceptions of the agent's personality.

The experiment was conducted on LINE, a widely used instant messaging service in Taiwan. To achieve this, we created a LINE bot that could change both its avatar and the name of its responses. Additionally, we utilized the LINE Front-end Framework to gather user feedback during each experimental condition. This allowed participants to complete the experiment using only the LINE app. The source code for the platform can be found at GitHub[1].

### 3.3 Natural Language Generation Model Design

Dialog agents utilize the GPT-3 (text-davinci-001) to generate responses. The prompts used for generated text include the agent's personality and the previous three dialogs for multi-turn conversation. Examples of input and agent settings can be found in Table 1 and supplementary materials. All text input is in English. If users input text in other languages, it undergoes pre-processing through Google Translate before being fed into the GPT-3 or dialog analyzers.

### 3.4 Data Collection

*3.4.1 Chat Experience.* We translated the Chatbot Usability Questionnaire (CUQ) [13] into Chinese to measure users' chat experience toward each agent. The purpose of the CUQ is to assess the participant's perception of their experience using chatbots. To do this, it presents 16 statements, such as 'The chatbot understood me well' or 'Chatbot responses were irrelevant', which the participant must rate on a scale of 1 (strongly disagree) to 5 (strongly agree) (Figure 2c). The

---

[1]Anonymous: https://example.com

Cronbach's Alpha of the collected CUQ data in the experiment is α = .95. Please see Appendix B for all CUQ questions we used.

To represent the chat experience as a reflective indicator in the conceptual model, we only use questions 1 - 12 in the CUQ for analysis. Questions 13 to 16 measure the application interface and error handling rather than the dialogue content. Consequently, these questions (13-16) exhibit unacceptable loading [12] when analysis.

*3.4.2 Affinity.* We evaluate affinity to represent users' attitude toward the dialog agents. First, users rated their affinity on a scale of 1 (lowest affinity) to 7 (highest affinity) based on the name and avatar of the dialog agent. After finishing the chat, they re-rated the affinity score of the agent based on the name, avatar, and chat history (Figure 2b).

*3.4.3 Intention of use.* To measure users' intention to use the AI dialog system, we employed the Technology Acceptance Model (TAM) scale translated by previous research [19]. This translated TAM scale could evaluate users' acceptance of NLG chatbots based on their intention to use, perceived ease of use, and perceived usefulness. Even though all three aforementioned TAM indicators can significantly predict CUQ, we have opted to analyze the "intention to use" factor only to simplify the conceptual model [12]. All Cronbach's Alpha of the collected TAM indicators are above α > 0.9.

## 3.5 Agent's Character and User's Perceptions

Our conceptual model uses five Big Five personality level dimensions to construct a formative indicator to represent the agent's character, controlled by different prompts. To quantify the personality designs, each big 5 dimensions is encoded as a dummy variable based on the assigned personality. For example, if the agent is extroverted, they are assigned a label of 1 within the extroversion dimension. Conversely, if the agent is introverted, their label within the extroversion dimension is set to -1. Moreover, if the agent's personality is openness, which is unrelated to extroversion, they receive a label of 0 within the extroversion dimension.

To ensure that our prompt successfully manipulates the agent's character, we ask the user to provide ratings on the dialog agent personalities from 1 (not related to this personality at all) to 10 (certainly related to this personality) for the five dimensions of the Big Five personality traits after completing other assessments (Figure 2d).

## 3.6 Dialog Indicators from Dialog Analyzers

We use NLP-model-based Dialog Analyzers in Python 3.7.10, employing the transformers library version 4.25.1. Outputs from the Coherence and Emotion analyzers will be combined and serve as a formative indicator in our conceptual model.

Please see the example at Figure 3 regarding how we employ dialog analyzers to generate coherence and emotion indicators. When collecting outputs from DialogRPT analyzers, we evaluate the dialog by inputting both user input and agent reply, as well as the agent reply and follow-up input from users. When using the emotion analyzers, we collect emotions from both the user's input and the agent's reply.

*3.6.1 Coherence.* We utilized the microsoft/DialogRPT-human-vs-rand analyzer [11] to represent the coherence of the conversation. This analyzer outputs a score between 0 and 1, indicating the probability of the response being relevant to the input. Developed by Microsoft Research's NLP Group, the DialogRPT analyzers were trained on the extensive Pushshift Reddit Dataset, which encompasses 100+ million pieces of human feedback data.

*3.6.2 Emotion.* We employed the cardiffnlp/twitter-roberta-base-emotion-multilabel-latest analyzer [3] to predict the emotions of the dialogue. The analyzer is designed to predict likelihood scores within the range of 0 to 1, indicating the level of correlation between the input provided and eleven distinct emotions. These emotions encompass anger, anticipation, disgust, fear, joy, love, optimism, pessimism, sadness, surprise, and trust. Each emotion label is independently generated without undergoing softmax-based weighting. The analyzer was fine-tuned by the NLP Research Group at Cardiff University using the TweetEval benchmark[1].

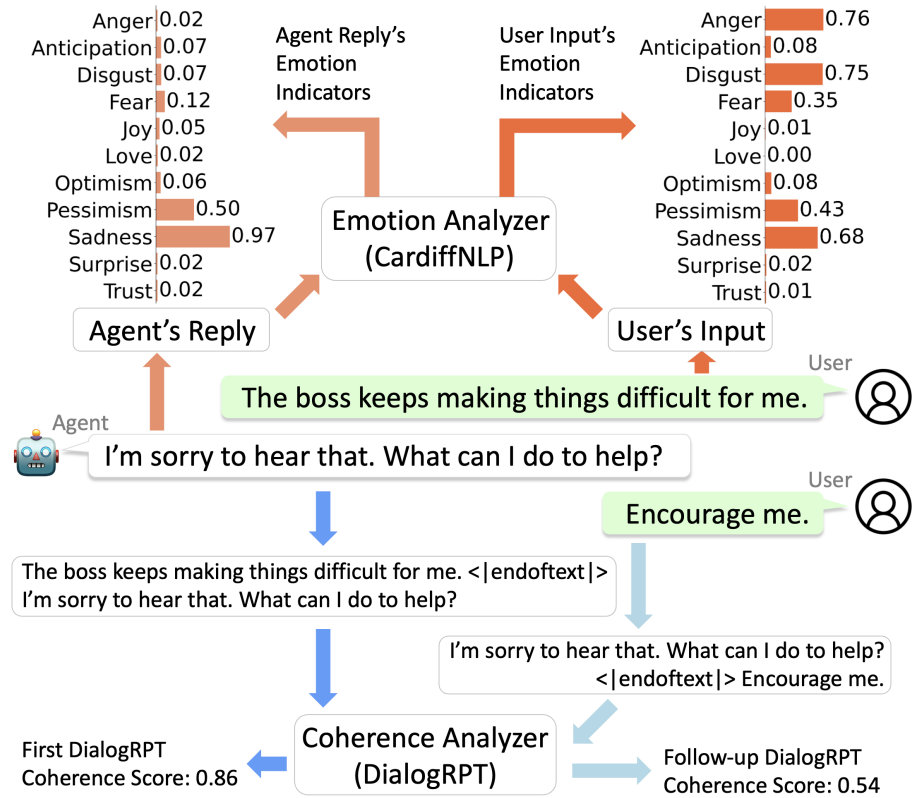

Fig. 3. Example of how to select dialog content and generate Coherence and Emotion Indicators by Dialog Analyzers.

### 3.7 Outputs from more dialog analyzers

To assess H9, which examines whether adding additional dialog indicators can lead to improved prediction accuracy, we integrated more outputs from four additional dialog analyzers combined with the existing Coherence and Emotion outputs. All these outputs served as another formative indicator. Then, we will compare the adjusted R-square and SRMR to see if more dialog indicators can reach higher accuracy. Additionally, we will further evaluate the effectiveness of each dialog indicator by separating them for further SEM analysis and doing the XGBoost Feature Importance Ranking.

The following sections are introductions to the other four analyzers we use:

*3.7.1 Human-Like.* The microsoft/DialogRPT-human-vs-machine analyzer[11] outputs a score between 0 and 1, indicating the likelihood that a given response was authored by a human rather than generated by a machine.

*3.7.2 Updown.* The microsoft/DialogRPT-updown analyzer[11] can estimate a score ranging from 0 to 1, which signifies the probability that the response will receive upvotes on Reddit. This score serves as an indicator of the response's impact. For instance, informative responses may receive upvotes, leading to a higher updown score, while inappropriate content may receive downvotes, resulting in a lower updown score.

*3.7.3 Width.* The microsoft/DialogRPT-width analyzer[11] can predict a score between 0 and 1, representing the likelihood that the response will elicit direct replies on Reddit. For instance, if the response asks a question, it will likely get answers, resulting in a higher width score.

*3.7.4 Depth.* The microsoft/DialogRPT-depth analyzer[11] can generate a score between 0 and 1, indicating the likelihood that the response will initiate longer follow-up discussions. Responses likely to trigger extensive conversations about a topic will have a higher depth score.

## 3.8 Data Analysis

In our study, 2400 samples were gathered, involving 120 participants interacting with 20 agents. The statistical analyses were conducted using SmartPLS 4 using Partial Least Squares Structural Equation Modeling (PLS-SEM). PLS-SEM, a second-generation multivariate analysis technique useful for examining relationships between latent independent and dependent variables[12]. Furthermore, for Correlation, t-test, and ANOVA, we utilize the SPSS 23. For XGBoost, we use the Xgboost library version 1.7.6 in Python 3.7.10.

First, we calculated the correlation between users' perceptions of the agent's character and the actual character of the agent. Subsequently, we evaluated the chat experience ratings based on various character designs for chat agents through ANOVA analysis.

Then, we used PLS-SEM to verify the construct validity of the proposed conceptual model. We evaluated the reliability, internal consistency, and convergent and discriminant validity through the confirmatory factor analysis.

After that, we used PLS-SEM to examine structural relationships among latent variables in the proposed model. The maximum likelihood method and bootstrapping resampling were employed to measure the statistical significance of the model's path coefficients and the mediation effect among variables.

Additionally, given our keen focus on the dialog indicators, we will analyze them separately to see their effectiveness in predicting Chat Experience. We will also explore variations in their predictive performance when using Emotion or Coherence individually and when integrating outputs with additional dialog indicators.

Finally, by utilizing the XGBoost algorithm [6], we computed the feature importance scores of all collected dialog indicators, allowing us to discuss which indicator has the most impact on the chat experience. XGBoost is an advanced algorithm that builds upon gradient-boosting decision trees, displaying expertise in crafting boosted trees through parallel processing. In addition to optimizing the objective function, XGBoost computes importance scores for each feature.

## 4 RESULTS

### 4.1 User perceptions of the agent's character versus its true character

Correlation results in Table 2 indicated that users' perception of the agent's character positively correlates with the actual character design. In other words, our prompt manipulation is successful. When the Agent is designed with a certain personality, users tend to consider the agent to have that personality. (H2 accepted)

Table 2. User perceptions of the agent's character versus its true character. Note. $^{***}p < .001$.

| Users' rating on agents' Big Five personality traits | Agents' designed Big Five personality traits in the same dimension |
|---|---|
| Agreeableness | .256*** |
| Extroversion | .089*** |
| Conscientiousness | .162*** |
| Neuroticism | .126*** |
| Openness | .142*** |

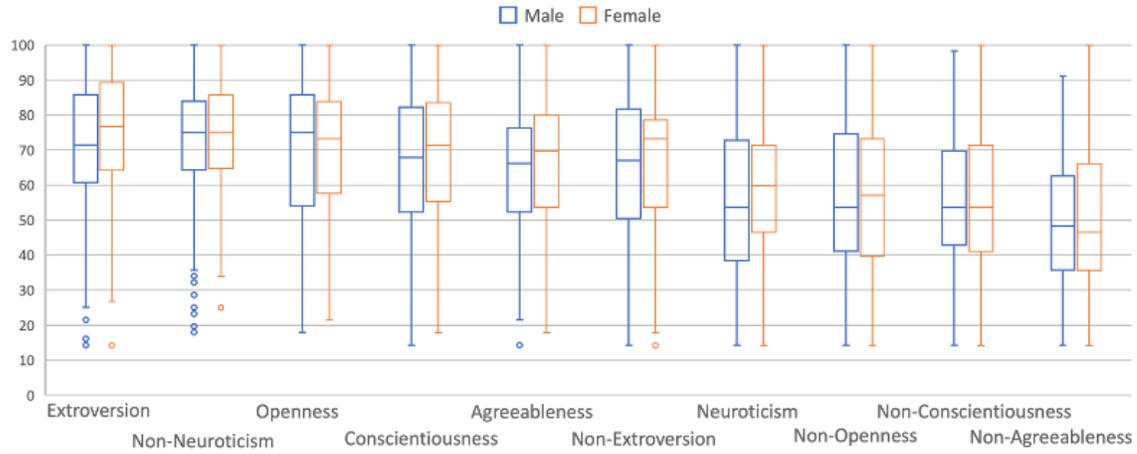

Fig. 4. The difference of chat experience score between different agents' Big Five personality and gender

## 4.2 Comparison between agents with high and low chat experience score

A two-way ANOVA was conducted on the Chat Experience Score for 20 prompt designs: 10 (Big Five personality and their opposites) x 2 (genders). The interaction between personality and gender was not significant ($F(9, 2380) = .288$, $p = .978$). Moreover, the main effects analysis revealed that different agents' characters ($F = 42.2$, $p < .001$, $\eta^2 = .137$) and gender ($F = 6.351$, $p = .012$, $\eta^2 = .002$) had a significant effect on chat experience (H1 accepted).

The results in Figure 4 indicated that different prompt designs influence the Chat Experience Score. Specifically, agents with personality traits such as Extroversion, Non-Neuroticism, Openness, Conscientiousness, and Agreeableness tend to receive higher chat experience scores ($t(2398) = 15.97$, $p < .001$, $Cohen's\ d = 0.652$).

## 4.3 Construct validity of the proposed model

The adjusted R square ($_{adj}R^2$) of the chat experience score on our proposed model is 0.536. In the estimated model, the Standardized Root Mean Square Residual (SRMR) is 0.04, and the Normed Fit Index (NFI) is 0.942. Moreover, the post-hoc statistical power analysis indicated the statistical power is 1.0, which means the model fits well [12, 15]. The item factor loadings from Chat Experience Score and User's Intention are above 0.7. The average variance extracted (AVE) of all latent variables are above 0.6. All latent variables' composite reliability (CR) and Cronbach's alpha (CA) are above 0.9. These results indicate that all latent variables showed good reliability and internal consistency. Due to the word limitations, detailed reliability and validity results are presented in Appendix C.

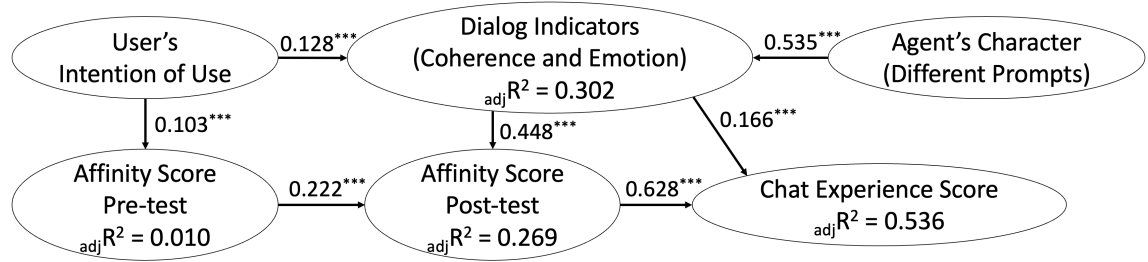

Fig. 5. The PLS-SEM result of the proposed model, indicated the relationship between User's Intention, Dialog Analyzers, Agent's Personality by Different Prompts, User's affinity score on agents, and the Chat Experience Score. Note. $^{***}p < .001$.

### 4.4 Structural relationships among latent variables

The PLS-SEM was used to examine the model of the structural relationships among the latent variables of the proposed model. The result is shown in Figure 5. Paths with no statistical significance ($p > .05$) were ignored.

The result indicated that the chat experience score ($_{adj}R^2 = .536$) could be predicted by post-test affinity score (path coefficient = .628, $p < .001$) and dialog indicators (path coefficient = .166, $p < .001$). Furthermore, the post-test affinity score ($_{adj}R^2 = .269$) can be predicted by the pre-test affinity Score (path coefficient = .222, $p < .001$) and dialog indicators (path coefficient = .448, $p < .001$). Moreover, the dialog indicators ($_{adj}R^2 = .302$) can be predicted by the agent's character by different prompts (path coefficient = .535, $p < .001$) and the user's intention of use (path coefficient = .127, $p < .001$).

To check whether dialog indicators (DI) and affinity score (PreAff and PostAff) mediated the relationship between the user's intention of use (UIU), agents' character by different prompts (Prompt), and the chat experience score (CES), a bootstrapping method [24] with a resample of 5,000 (95% percentile confidence level) was performed. The results were determined according to suggestions by [31] and are shown in Table 3. Moreover, we used the same method to check whether the PostAff mediated the relationship between the PreAff and CES, as well as DI and CES. The results are also shown in Table 3.

The result in Figure 5 and Table 3 highlights the importance of dialog indicators in predicting chat experience. To elaborate, Dialog indicators were found to fully mediate the influence of the agent's character (different prompts) and the user's intention on the user's chat experience. Additionally, dialog indicators were partially mediated by post-test affinity scores.

Table 3. Mediation Analysis Results between Chat Experience Score (CES), User's Intention of Use (UIU), Dialog Indicators (DI), post-test affinity Score (PostLS), and pre-test affinity Score (PreLS), and Agents' Character by Different Prompts (Prompt). Note $^{***}$ $p < .001$.

| Tested Relationship | Direct Model | Indirect Model | Results |
|---|---|---|---|
| Prompt → DI → CES | 0.007 | 0.088*** | Indirect-only Mediation |
| Prompt → DI → PostAff → CES | 0.007 | 0.151*** | Indirect-only Mediation |
| DI → PostAff → CES | 0.165*** | 0.282*** | Complementary Mediation |
| UIU → DI → CES | 0.023 | 0.021*** | Indirect-only Mediation |
| UIU → DI → PostAff → CES | 0.023 | 0.036*** | Indirect-only Mediation |
| UIU → PreAff → PostAff → CES | 0.023 | 0.014*** | Indirect-only Mediation |
| PreAff → PostAff → CES | 0.018 | 0.139*** | Indirect-only Mediation |

### 4.5 Adding more dialog indicators

Table 4 shows the predicting power of different dialog indicators. When using only Dialog's Coherence and Emotion as dialog indicators, the chat experience's adjusted $R^2$ is .243. When solely utilizing Dialog's Coherence, the adjusted $R^2$ is .168. Similarly, when exclusively relying on Dialog's Emotion, the adjusted $R^2$ is .203. Furthermore, when only using outputs from other DialogRPT analyzers from subsection 3.7, the adjusted $R^2$ is .134. Finally, when incorporating additional indicators from subsection 3.7 along with coherence and emotion indicators, the adjusted $R^2$ increases from .243 to .249.

Table 4. The predictive power of different features selection set. Note. *** $p < .001$.

| Feature Selection | $R^2$ | Path Coefficient |
|---|---|---|
| Emotion | 0.203 | 0.45*** |
| Coherence | 0.168 | 0.41*** |
| Other 4 DialogRPT Analyzers' Outputs | 0.134 | 0.37*** |
| Coherence + Emotion | 0.243 | 0.49*** |
| Coherence + Emotion + Other 4 DialogRPT Analyzers' Outputs | 0.249 | 0.5*** |

We incorporated outputs from other dialog analyzers into our conceptual model alongside the existing Coherence and Emotion as a new formative dialog indicator. Then, we run the PLS-SEM analysis again. The findings revealed that the significant pathways remained consistent with those obtained when considering only Coherence and Emotion, as depicted in Figure 5. Moreover, the $R^2$ increased from 0.536 to 0.537 and the SRMR improved from 0.04 to 0.038.

These results indicate that including additional dialog indicators can enhance predictive performance on the chat experience, supporting the acceptance of H9.

We train an XGBoost model on predicting chat experience with indicators from all dialog analyzers to obtain the feature importance weights. The result is in Figure 6. The result revealed that the most effective predictors primarily revolve around coherence and emotion, rather than relying on other additional analyzers. The top ten indicators for predicting the chat experience are the following: Agent's Optimism, DialogRPT-human-vs-rand, User's Fear, User's Anger, Agent's Disgust, User's Sadness, User's Love, User's Hate, User's pessimism, and User's Surprise.

## 5 DISCUSSION

### 5.1 The effect of different Prompts

Our research employs prompt engineering to shape the personality traits of dialogue agents, causing them to exhibit distinct character across the Big 5 personality dimensions. A correlation analysis highlights the success of these adjustments, as they result in perceptible shifts in how users perceive the personality of the dialogue agent. Furthermore, ANOVA analysis reveals that dialogue agents who exhibit any of the following characteristics are likely to receive higher chat experience scores: Extroversion, Low Neuroticism, Openness, Conscientiousness, and Agreeableness.

How we designed the prompt of dialogue agents significantly impacts how users perceive their chat experience and affinity toward agents within the dialog system. Moreover, dialog indicators generated by NLP-based dialog analyzers can explain these effects. This suggests the potential for creating an adaptive dialogue system that utilizes dialog indicators to predict real-time chat experiences and optimize the conversation accordingly by adjusting prompts to generate responses that lead to appropriate dialog indicators. For example, an adaptive dialogue system can actively assess the coherence and emotion of the conversation and, when necessary, employ different prompts to maintain these indicators at satisfactory levels, ultimately enhancing the chat experience.

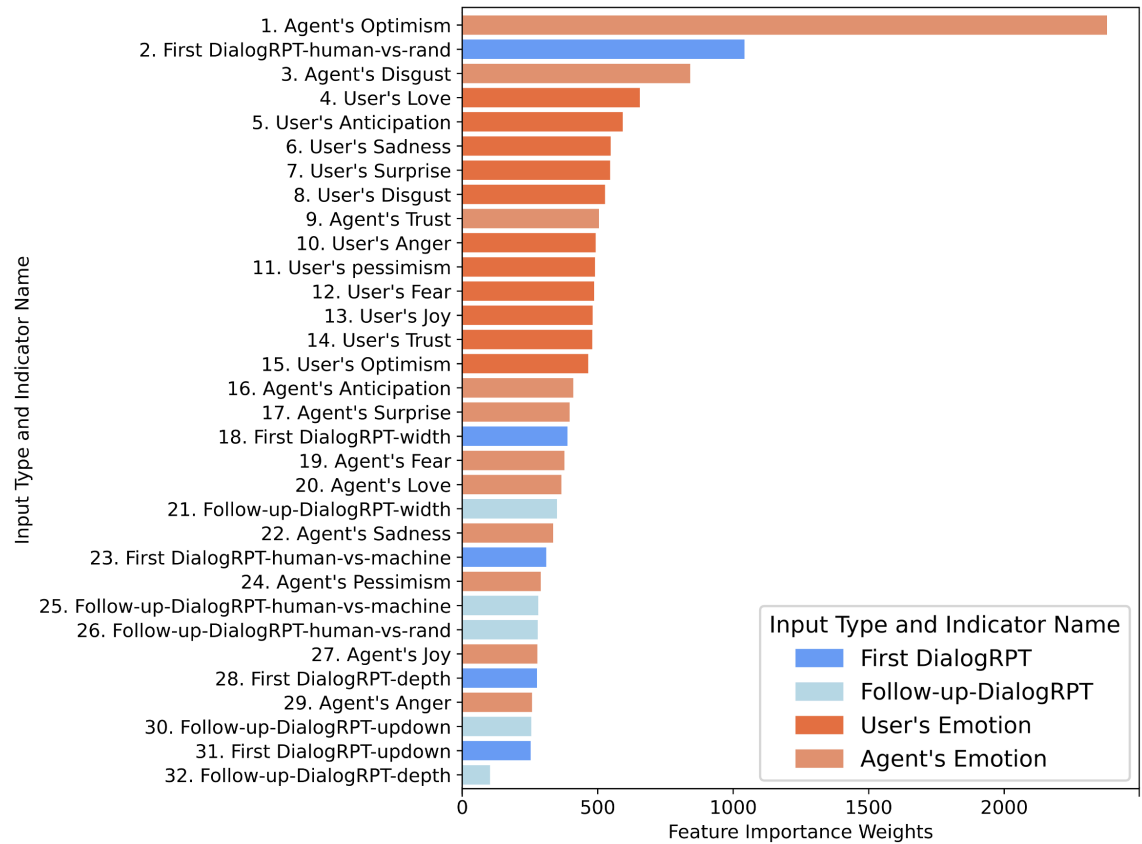

Fig. 6. The feature important result of dialog indicators by the XGBoost algorithm.

## 5.2 Effectiveness of the proposed model

Our conceptual model, as illustrated in Figure 5, demonstrated a good fit (SRMR = .041) and significant predictive power for chat experience scores ($_{adj}R^2$ = .536). The result underscores the crucial role of dialog indicators, such as coherence and emotion, in predicting chat experience. These indicators fully explained how the agent's character (prompts) and the user's intention to use the system affected the chat experience. Moreover, the effect of dialog indicators was partly explained by post-test affinity scores. Lastly, our research revealed that our conceptual model's robustness stayed consistent when incorporating more dialog indicators. Therefore, future researchers and developers may consider employing our conceptual model as an effective framework to evaluate and enhance the chat experience.

## 5.3 Using dialog indicators to predict chat experience

Our findings demonstrate the effectiveness of dialog indicators in predicting chat experiences ($R^2$ = 0.243). Furthermore, our results suggest that incorporating additional dialog indicators could further improve prediction performance, increasing $R^2$ to 0.249. However, it is worth noting that the increase in predictive performance achieved by adding extra indicators is limited. Moreover, the Feature Importance analysis highlights that the weight of Coherence ($R^2$ = 0.168) and Emotion ($R^2$ = 0.203) in the conversation are more important than the additional dialog indicators ($R^2$ = 0.134) we added. This implies that although performance can be slightly improved by adding more dialog analyzers, it also complicates the interpretability of the predictions. An interesting question for future research is how to efficiently select the most effective dialog indicators and combine them to predict chat experience cost-effectively. Additionally, it is worth exploring finding a new dialog indicator independent of coherence and emotion while having strong predictive power on the chat experience.

Furthermore, while it's possible to predict dialog indicators based on the user's intention and agent's character by different prompts, the adjusted $R^2$ value linked to this relationship is merely 0.3. Moreover, the post-test affinity score can only partially explain the influence of dialog indicators on chat experience. This suggests that indicators from dialog analyzers may have greater complexity beyond these factors, and numerous unexplored variables could be explained or used to explain the output from dialog analyzers. This presents an intriguing avenue for future research.

### 5.4 The importance of users' affinity on agents

Our findings showed that users' pre-test affinity score strongly predicts the chat experience. Furthermore, users' post-test affinity score acted as a complementary mediator for dialog indicators, offering insights into the variance observed in these indicators. Therefore, we suggest that asking users how they feel about the dialog agents may be an effective way to estimate their overall chat experience.

## 6 LIMITATIONS

Firstly, it's important to acknowledge that the specific demographics of our participants may influence our findings. Furthermore, our study's limitations encompass the chat software and the controlled experimental environment we utilized, as user conversational patterns in the experiment may differ from real-world interactions. Additionally, we used the text-davinci-001 model, which may not represent the performance of the latest models, such as GPT-4. Despite these limitations, we made every effort to ensure that the user-AI chat experience closely resembled real-world conditions and recruited a diverse range of participants to enhance the robustness of our study.

## 7 CONCLUSION

In this paper, we introduced a conceptual model to explain the relationship between dialog indicators and various factors related to the chat experience, such as users' intentions, users' affinity toward agents, and the characters of the dialogue agents modified by different prompts. We assessed this model with 120 participants using PLS-SEM, achieving an R-squared ($R^2$) value of 0.535 and SRMR of 0.041. We then compared the importance of different dialog indicators in predicting chat experiences. Our findings highlight Emotion and Coherence as the most influential indicators. These results demonstrate that dialog indicators, particularly coherence and emotion, can effectively predict chat experiences, acting as full mediators for the influence of prompts and user intentions. Additionally, users' affinity toward agents can partially explain these predictions. Finally, we discuss these results and provide suggestions for future research.

### ACKNOWLEDGMENTS

The study was approved by the Institutional Review Board of Anonymous University with protocol code (XXXXXXXX). This work was supported by Anonymous under Grants XXXX.